\documentclass[fleqn,usenatbib]{mnras}

\usepackage[english]{babel}
\usepackage{mathtools}

\DeclarePairedDelimiter\floor{\lfloor}{\rfloor}

\usepackage{amsmath}
\usepackage{amssymb}
\usepackage{graphicx}
\usepackage{physics}
\usepackage{subcaption}
\usepackage{graphicx}
\usepackage{xcolor}
\usepackage{verbatim}
\usepackage{lmodern}

\newcommand{\Mmed}{M_{\rm med}}
\newcommand{\Msun}{{\rm M}_{_\odot}}
\newcommand{\AN}{{\cal A}_{_N}}
\newcommand{\AS}{{\cal A}_{_7}}
\newcommand{\abfn}{\boldsymbol{a}_n}
\newcommand{\abfnp}{\boldsymbol{a}_{n'}}

\newcommand{\alfLAW}{\alpha_{\rm law}}
\newcommand{\alfROT}{\alpha_{\rm rot}}
\newcommand{\alfSEG}{\alpha_{\rm seg}}
\newcommand{\DtMMO}{\Delta t_{_{\rm MMO}}}
\newcommand{\DND}{\Delta\!N_{_{\cal D}}}
\newcommand{\Ekin}{E_{\rm kin}}
\newcommand{\ebfhatz}{\hat{\boldsymbol e}_z}
\newcommand{\etD}{\eta_{_{\cal D}}}
\newcommand{\fCDm}{f_{_{{\cal CD}m}}}
\newcommand{\fM}{f_M}
\newcommand{\fR}{f_R}
\newcommand{\fTp}{f_{_{\rm T^+}}}
\newcommand{\Iz}{I_z}
\newcommand{\lgt}{\log_{10}}
\newcommand{\Mmd}{M_{\rm med}}
\newcommand{\Mn}{M_n}
\newcommand{\Mtot}{M_{\rm tot}}
\newcommand{\mmax}{m_{\rm max}}
\newcommand{\muD}{\mu_{_{\cal D}}}
\newcommand{\muell}{\mu_\ell}
\newcommand{\muN}{\mu_{_{N}}}
\newcommand{\NCDm}{{\cal N}_{_{{\cal CD}m}}}
\newcommand{\ND}{N_{_{\cal D}}}
\newcommand{\Om}{{\cal O}_m}
\newcommand{\PDN}{{\cal P}_{_{{\cal D}N}}}
\newcommand{\QCD}{{\cal Q}_{_{\cal CD}}}
\newcommand{\rbfn}{\boldsymbol{r}_n}
\newcommand{\rbfnp}{\boldsymbol{r}_{n'}}
\newcommand{\SCNfp}{{\cal S}_{_{{\cal C}N4^+}}}
\newcommand{\SCNm}{{\cal S}_{_{{\cal C}Nm}}}
\newcommand{\sigD}{\sigma_{_{\cal D}}}
\newcommand{\sigell}{\sigma_\ell}
\newcommand{\sigISO}{\sigma_{\rm iso}}
\newcommand{\sigN}{\sigma_{_N}}
\newcommand{\vbfn}{\boldsymbol{v}_n}
\newcommand{\vbfnp}{\boldsymbol{v}_{n'}}
\newcommand{\vbfISO}{\boldsymbol{v}_{\rm iso}}
\newcommand{\vbfISOn}{\boldsymbol{v}_{{\rm iso:}n}}
\newcommand{\vbfROT}{\boldsymbol{v}_{\rm rot}}
\newcommand{\vbfROTn}{\boldsymbol{v}_{{\rm rot:}n}}

\begin{document}
\title[The formation of multiples in small-{\it N} subclusters]{The formation of multiples in small-{\it N} subclusters}
\author[H. E. Ambrose, A. P. Whitworth]
{Hannah E. Ambrose$^1$\thanks{E-mail: ambrosehe@cardiff.ac.uk}, A. P. Whitworth$^1$\\
$^1$CHART, School of Physics and Astronomy, Cardiff University, Cardiff CF24 3AA, UK}
\date{Accepted XXX. Received YYY; in original form ZZZ}
\pubyear{2024}
\label{firstpage}
\pagerange{\pageref{firstpage}--\pageref{lastpage}}
\maketitle

\begin{abstract}
We explore the relative percentages of binary systems and higher-order multiples that are formed by pure stellar dynamics, within a small subcluster of $N$ stars. The subcluster is intended to represent the fragmentation products of a single isolated core, after most of the residual gas of the natal core has dispersed. Initially the stars have random positions, and masses drawn from a log-normal distribution. For low-mass cores spawning multiple systems with Sun-like primaries, the best fit to the observed percentages of singles, binaries, triples and higher-order systems is obtained if a typical core spawns on average between $N=$ 4.3 and 5.2 stars, specifically a distribution of $N$ with mean $\muN\sim4.8$ and standard deviation $\sigN\sim2.4$. This fit is obtained when $\sim50\%$ of the subcluster's internal kinetic energy is invested in ordered rotation and $\sim50\%$ in isotropic Maxwellian velocities. There is little dependence on other factors, for example mass segregation or the rotation law. Whilst such high values of $N$ are at variance with the lower values often quoted (i.e. $N\!=\!1\;{\rm or}\;2$), very similar values ($N\!=\!4.3\pm0.4$ and $N\!=\!4.5\pm1.9$) have been derived previously by completely independent routes, and seem inescapable when the observed distribution of multiplicities is taken into account.
\end{abstract}


\section{Introduction}\label{intro}

Most field stars more massive than the Sun are not single \citep{2015MNRAS.448.1761W, 2023ASPC..534..275O}. They exist in multiple systems, i.e.  gravitationally bound groups in which the individual stars are on more-or-less constant and stable orbits about one another. The most common multiple systems are binaries: two stars orbiting their mutual centre of mass on elliptical orbits. However, there is an increasing recognition that there are also many higher-order multiple systems in the field (and even more in regions of recent star formation , \citealt{2013ApJ...768..110C}). Stable, long-lived triple systems comprise two stars in a relatively close binary system with a third star on a significantly wider orbit about the binary. Quadruple systems come in two variants: `2+2' quadruples comprise two relatively close binary systems on a wide orbit about one another; and `planetary' quadruples comprise a triple system (as define above) with a fourth star on an even wider orbit about the triple. The highest-order multiples known are septuples. An example of a septuple is 65 UMa (11551+4629), in which a central binary is orbited not only by three companions on planetary orbits, but a distant binary pair as well (see \citealt{2021Univ....7..352T}).

A young subcluster comprising $N$ stars may evolve into $\AN$ possible end states, where $\AN$ increases rapidly with $N$. For $N\leq 10$:
\begin{eqnarray}\nonumber\
    \AN\!\!&\!\!\!=\!\!&\!\!\sum_{n=2}^{N}\!\left\{ 1+\sum_{n'=2}^{n}\!\left\{\vphantom{\sum_{n''=2}^{n'-1}}\!\floor*{\frac{N\!-\!n}{n'}}\,+\,H(N\!-\!(n+n'+2))\;\;\right.\right. \\\label{EQN:decay}
   \!\!&\!\!\!\!\!&\!\!\hspace{3.0cm}\left.\left.\sum_{n''=2}^{n'-1}\floor*{\frac{N-(n+n')}{n''}}\right\}\right\}.
\end{eqnarray}
Here $\floor*{\;}$ is the floor operator, and $H$ is the Heaviside function. Thus a subcluster with $N=7$ members can evolve into $\AS=14$ different end states, as enumerated in Section \ref{SEC:UniversalN}. A derivation and explanation of Equation \ref{EQN:decay} is given in Appendix \ref{appendA1}.

\subsection{Observational perspective}

Recent advances in observational technology and capability have lead to the discovery of additional members in known star systems, and thereby greatly increased the inventory of higher-order multiples in the solar neighbourhood. When \citet{1991A&A...248..485D} studied the multiplicity of solar-type stars in a 22-pc sample, they found that only 5\% hosted triple or higher order systems. For a 25-pc sample, \citet{2010ApJS..190....1R} found this percentage to be 13\%, and it increased to 17\% when \citet{2021Univ....7..352T} and \citet{2021AJ....161..134H} studied the sample in 2021.

The \citet{2021Univ....7..352T} statistics for nearby systems with solar-type primaries, in the field (hereafter the T21 sample),  are
\begin{eqnarray}\label{EQN:Tokstat}
  \rm{S:B:T:Q^+}&=&54:29:12:5\,.
\end{eqnarray}
Here S is the percentage of single stars, B the percentage of binaries, T the percentage of triples, and Q$^+$ the percentage of quadruples plus higher-order systems (i.e. quintuples, sextuples and septuples).

Thus, in this sample, the fraction of stars that are single is
\begin{eqnarray}
f_{_{\rm S}}&=&\frac{\rm S}{\rm S+2B+3T+4Q^+}\;\;\,\simeq\;\;\,32.1\%\,;
\end{eqnarray}
the fraction of stars that is in binaries is 
\begin{eqnarray}
f_{_{\rm B}}&=&\frac{\rm 2B}{\rm S+2B+3T+4Q^+}\;\;\,\simeq\,\;\;34.6\%\,;
\end{eqnarray}
and the fraction that is in triples or higher order systems is
\begin{eqnarray}\label{EQN:fTp.01}
\fTp&=&\frac{\rm 3T+4Q^+}{\rm S+2B+3T+4Q^+}\;\;\,\simeq\,\;\;33.3\%\,.
\end{eqnarray}

These percentages should be qualified with two caveats. First, we have assumed that all the systems contributing to Q$^+$ are quadruples. If account were taken of the systems that are quintuples, sextuples and septuples, $\fTp$ would increase slightly, at the expense of $f_{_{\rm S}}$ and $f_{_{\rm B}}$. Second, although these percentages are uncertain, due to selection effects and  observational bias, the expectation must be that in the future they will shift towards higher-orders, as additional members of existing systems are discovered. This will reduce $f_{_{\rm S}}$, and probably increase both $f_{_{\rm B}}$ and $f_{_{\rm T^+}}$.

Moreover, we should also note that, by construction, the stars in the T21 sample have masses $M\lesssim{\rm M}_{_\odot}$. Samples with higher-mass primaries tend to have even higher multiplicities than the T21 sample. For example,  in the survey by \cite{2017IAUS..329..110S} essentially all of the high-mass stars are members of higher-order systems. The systems we are considering here are presumed to be the product of low-mass cores of the sort that define the peak of the core mass function and are observed in nearby low-mass star formation regions.

We should also be mindful that the multiplicities of stellar systems depend on their age. \citet{1993A&A...278...81R} show that young populations tend to have higher multiplicities than their older counterparts. Indeed, multiplicity begins to decline even in the protostellar phase. \citealt{2013ApJ...768..110C} show that the average multiplicity decreases as one progresses from Class 0, through Class I, to Class II protostars.

\subsection{Theoretical perspective}\label{sec:introtheory}

We shall assume that usually an observed multiple system with a solar-mass primary has formed from a single prestellar core. There are then two main mechanisms involved: (i) dynamical core collapse and fragmentation; and (ii) disk fragmentation. These two mechanisms are not mutually exclusive, rather they tend to be sequential. \footnote{We note that in the higher-mass cores that form more massive stars a third mechanism needs to be considered. These higher-mass cores are often fed by filamentary accretion streams that are sufficiently massive to form multiple systems by filament fragmentation. These multiples systems then fall into the monolithic cluster that is forming in the central hub, where they interact with other stars and systems. This is not the scenario we are considering here.}

In the standard paradigm for star formation, molecular clouds are highly turbulent, and in some regions the turbulence produces convergent flows of sufficient strength to produce self-gravitating sheets, filaments and cores.

Unless it is rapidly dispersed by tidal forces or shear, a self-gravitating core collapses. And, unless it is very spherically symmetric and non-rotating, it is likely to fragment dynamically to produce a small subcluster of stars \citep[e.g.][]{2003MNRAS.340...91C, 2004A&A...414..633G, 2004ApJ...600..769F, HennebellePetal2004, 2004A&A...414..633G, GoodwinSetal2004b, HubberDWhitworthA2005, 2007MNRAS.382L..30S, 2010ApJ...725.1485O, WalchSetal2012, LomaxOetal2014, LomaxOetal2015a, LomaxOetal2015b, 2016PASA...33....4L, RohdePetal2021, WhitworthAetal2024}. This is `dynamical core collapse and fragmentation'.

The resulting stars are usually attended by accretion discs, comprising material that has too much specific angular momentum to fall directly onto the central star. If such an accretion disc becomes sufficiently massive, extended and cold, it fragments to produce a secondary star or stars, in orbit around the primary \citep[e.g.][]{1989ApJ...347..959A, 1992Natur.359..207C, 1994MNRAS.271..999B, 1995MNRAS.277..705T, 1995MNRAS.277..727W, 1998MNRAS.297..435B, 1998MNRAS.300.1189B, 1998MNRAS.300.1205W, 1998MNRAS.300.1214W, 2007MNRAS.382L..30S, 2008A&A...480..879S, 2009MNRAS.392..413S, 2009MNRAS.400...13W, 2009MNRAS.400.1563S, 2010ApJ...717..577T, 2010MNRAS.402.2253W, 2010ApJ...708.1585K, 2011ApJ...730...32S, 2012MNRAS.427.1182S, LomaxOetal2015a, LomaxOetal2015b, 2016PASA...33....3W, 2023MNRAS.519.2578D}. This is `disc fragmentation'.

Together, these mechanisms have been shown to produce young multiple systems of as many as seven stars \citep{2016PASA...33....4L}.

\subsection{Previous related theoretical studies}

\citet{1993MNRAS.262..800M} have introduced the concept of dynamical biasing, which is the tendency for the more massive stars in a small-$N$ subcluster to form a binary, and for the less massive ones to be ejected, thereby making the binary more tightly bound. In a second paper, \citet{1995MNRAS.275..671M} show that, if the stars in a small-$N$ subcluster are attended by discs, the increased dissipation during close encounters reduces the effectiveness of dynamical biasing -- i.e. the lower-mass stars have a better chance of ending up in a binary. It also increases the tendency to produce higher-order multiples. \cite{2023A&A...674A.196K} have shown that the drag from the gas of the natal cloud can drive the inspiral of forming stars, bringing binaries closer together. This would harden them against dissolution and increase the proportion of higher-order multiples.

\cite{1998A&A...339...95S} have explored multiplicity in a cold, non-rotating, spherical distribution of $N=$ 3, 4, or 5 stars. The stars have masses drawn from a Miller-Scalo distribution, and are evolved for 1000 crossing times, using an $N$-body code. At the end the multiplicities are
\begin{eqnarray}\begin{array}{lcll}
\rm{S:B:T}&=&47:47:6, \hspace{0.7cm}&\rm{for}\;\,N=3;\\
\rm{S:B:T:Q }&=&63:29:6:1,&\rm{for}\;\,N=4;\\
\rm{S:B:T:Q^+}&=&70:19:9:1,&\rm{for}\;\,N=5.\\
\end{array}
\end{eqnarray}
Thus, increasing $N$ increases the percentage of singles and reduces the percentage of binaries.

\citet{2013MNRAS.432.3534H} have shown using purely statistical arguments that, if the shape of the stellar Initial Mass Function (IMF) is inherited from the shape of the prestellar Core Mass Function (CMF), then the observed increase in binary frequency with primary mass \citep[e.g.][]{2023ASPC..534..275O} requires that a prestellar core must typically spawn a subcluster of 4 to 5 stars.

\subsection{Overview and plan}

In this work we extend the study of \cite{1998A&A...339...95S} to larger numbers of stars, and to more general initial conditions. Although circumstellar discs and disc fragmentation are likely to be important, particularly for forming lower mass stars and close orbits, we limit consideration here to subclusters formed purely by dynamical core collapse and fragmentation, and neglect discs. Consequently our numerical experiments start with single discless stars, and all multiple systems are formed subsequently by energy-conserving gravitational interactions between point masses. The role of discs will be explored in a later paper.

In Section \ref{Method} we describe our model for a small-$N$ subcluster, and the procedures used to follow its evolution and to identify the resulting multiple systems. In Section \ref{results} we analyse the results, and by comparing them with observations derive the best-fit parameters for a subcluster. In Section \ref{conclusion} we summarise the main conclusions. In Appendix \ref{appendA1} we derive Equation \ref{EQN:decay}, and in Appendix \ref{appendA2} we show how the results can be re-scaled to represent subclusters of different initial mass and/or linear extent. In Table 1 we list the parameters used in this paper.

\begin{table}
\label{TAB:Parameters}
\caption{Parameters and symbols}
\begin{center}\begin{tabular}{ll}\hline
$N$ {\sc and the Configuration Parameters} & \\
number of stars in subcluster & $N$ \\
standard deviation of $\lgt(M/\Msun)$ & $\sigell$ \\
percentage of kinetic energy in ordered rotation & $\alfROT$ \\
rotation law: solid-body={\sc sol}; Keplerian={\sc kep} & $\alfLAW$ \\
mass segregation option & $\alfSEG$ \\
Configuration, $[\sigell,\alfROT,\alfLAW,\alfSEG]$ & ${\cal C}$ \\\hline
{\sc Scaling Parameters} & \\
mean of $\lgt(m/\Msun)$ & $\muell$ \\
radius of subcluster & $R$ \\\hline
{\sc Numerical Parameters} & \\
adaptive integration timestep & $\Delta t$ \\
coefficient for integration timestep & $\gamma$ \\\hline
{\sc Analysis Parameters} & \\
percentage of singles ($m\!=\!1$) & S \\
percentage of binaries ($m\!=\!2$) & B \\
percentage of triples ($m\!=\!3$) & T \\
percentage of higher-order systems ($m\!\geq\!4$) & Q$^+$ \\
multiplicity of system & $m$ \\
maximum multiplicity considered & $\mmax$ \\
semi-major axis of orbit & $a$ \\
period of orbit & $P$ \\
eccentricity of orbit & $e$ \\
inclination of orbit & $\hat{\boldsymbol k}$ \\
time-interval for monitoring multiplicity & $\DtMMO$ \\
mean number of systems with multiplicity & \\
\hspace{0.65cm}$m$ formed by a single subcluster with & \\
\hspace{0.4cm}Configuration  $\,{\cal C}\,$ that contains $\,N\,$ stars & $\SCNm$ \\
probability that a core spawns $N$ stars & $\PDN$ \\
parameter for $\PDN$ (see Eqn. \ref{EQN:PDN.01}) & $\ND$ \\
parameter for  $\PDN$ (see Eqn. \ref{EQN:PDN.01}) & $\DND$ \\
normalisation coefficient for $\PDN$ (see Eqn. \ref{EQN:etD.01}) & $\etD$ \\
mean of $N$ for $\PDN$ (see Eqn. \ref{EQN:muD.01}) & $\muD$ \\
standard deviation of $N$ for $\PDN$ (see Eqn. \ref{EQN:sigmaN.01}) & $\sigD$ \\
$N$-distribution, $[\ND,\,\DND]$ & ${\cal D}$ \\
total number of systems with multiplicity $m$ & \\
\hspace{2.0cm}predicted for Configuration ${\cal C}$ & \\
\hspace{3.0cm}and $N$-distribution ${\cal D}$ & $\NCDm$ \\
observed number of systems with multiplicity $\!m$ & $\Om$ \\
quality of fit to observations with & \\
\hspace{1.25cm}Configuration ${\cal C}$ and $N$-distribution ${\cal D}$ & $\QCD$ \\
percentage of stars with $m\geq3$ & $\fTp$ \\\hline
{\sc General Variables and Functions} & \\
number of different end-states & \\
\hspace{2.0cm} for a subcluster of $N$ stars & $\AN$ \\ 
initial total kinetic energy of subcluster & $\Ekin$ \\
Heaviside Function & $H$ \\
random linear deviates on $[0,1]$& ${\cal L}_1\!,{\cal L}_2\!,{\cal L}_3$ \\
$\lgt(m/\Msun)$ (random Gaussian deviate) & $\ell$ \\
mass of star & $M$ \\
median mass & $\Mmd$ \\
total mass of subcluster & $\Mtot$ \\
dummy ID of star & $n,n',n''$ \\
position of star & $\boldsymbol{r}$ \\
 & $[r,\theta,\phi]$ \\
 & $[x,y,z]$ \\
velocity of star & $\boldsymbol{v}$ \\\hline
{\sc Scaling Parameters} & \\
factor to scale total mass of subcluster & $\fM$ \\
factor to scale radius of subcluster & $\fR$ \\\hline
\end{tabular}\end{center}
\end{table}

\section{Method} \label{Method}

\subsection{Initial conditions}

Five physical parameters are required to generate the initial conditions for a subcluster. The first is the number of stars in the subcluster, $N.$ We evolve subclusters with $3\leq N\leq 7$. $N$ can also equal 1 or 2, but we do not need to evolve these cases because their multiplicity statistics cannot be changed by pure internal dynamical evolution. Larger values of $N\;(>\!7)\,$ are not considered because it is prohibitive to keep track of all the different possible outcomes; such large $N$ are also likely to be rare.

The remaining four physical parameters, $(\sigell$, $\alfROT$, $\alfLAW$, $\alfSEG)\,$ are termed `Configuration Parameters', and regulate -- in a statistical sense -- the distribution of stellar masses and the initial distribution of stars in phase space. In the remainder of this section we define the four Configuration Parameters and explain how they are implemented in setting up the initial conditions. They are also listed in the first section of Table 1, and they are all dimensionless.

The individual stellar masses, $M$, are generated from a log-normal distribution. Specifically, $M=\Msun\,10^\ell$, where $\ell$ is a random deviate from a Gaussian distribution with mean $\muell$ and standard deviation $\sigell$. Here $\sigell$ is a Configuration Parameter, since it regulates the width of the mass distribution, i.e. the mean ratio between the most and least massive stars. We have explored values in the range $0.2\leq\sigell\leq 0.4$, but focus here on results obtained with $\sigell\!=\!0.3$.\footnote{Lower values of $\sigell$ result in too few binaries with low mass-ratios, $q$. Higher values result in an IMF that is too much broader than the CMF.} $\muell$ is {\it not} a Configuration Parameter, since the results can be re-scaled to any median mass, as explained in Appendix \ref{appendA2}. However, for the purpose of illustration we set $\muell=-0.6$, so that the median mass is $\Mmed\!=\!0.25\,\Msun$.

In the first instance, the stars are given random positions in a sphere of radius $R$. Specifically, the position of a star in spherical polars is $[r,\theta,\phi]=[{\cal L}_1^{1/3}R,\cos^{-1}(2{\cal L}_2\!-\!1),2\pi{\cal L}_3]$, where $[{\cal L}_1,{\cal L}_2,{\cal L}_3]$ are independent random linear deviates on the interval $[0,1]$. The spherical polar coordinates are then converted to Cartesian coordinates. $R$ is {\it not} a Configuration Parameter, since the results can be re-scaled to any subcluster radius, as explained in Appendix \ref{appendA2}. However, for the purpose of illustration we set $R=10^3\,{\rm AU}$. With $\muell=-0.6$ and $R=10^3\,{\rm AU}$, the  crossing time for the subcluster is $t_{\rm cross}\sim 0.07\,{\rm Myr}\,N^{-1/2}$.

The last three Configuration Parameters regulate the kinetic energy of the stars in the subcluster, its shape, and the spatial distribution of the stars. $\alfROT$ dictates the percentage of kinetic energy that is in ordered rotation, as opposed to random isotropic velocity dispersion. $\alfLAW$ dictates the rotation law: solid-body (Equation \ref{EQN:SOL}, SOL) or Keplerian (Equation \ref{EQN:KEP}, KEP). $\alfSEG$ determines whether the subcluster is mass-segregated at the outset. The order in which these parameters are invoked, when setting up the initial conditions, is dictated by computational considerations.

If $\alfSEG=1$, the stars are segregated by mass. In other words, the masses and positions are matched so that, for any pair of stars with IDs $n$ and $n'$, if $M_n<M_{n'}$, then $r_n>r_{n'}$. Otherwise $\alfSEG=0$ and the masses and positions are matched randomly. Once the masses and positions are matched, the centre of the Cartesian coordinate system is shifted to the centre of mass, and the initial self-gravitational potential energy,
\begin{eqnarray}
\Omega&=&-\,G\;\sum\limits_{n=1}^{n=N-1}\,\sum\limits_{n'=n+1}^{n'=N}\,\left\{\frac{M_n\,M_{n'}}{|\boldsymbol{r}_n-\boldsymbol{r}_{n'}|}\right\},
\end{eqnarray}
is computed.

At this juncture, the stars are given random isotropic velocities, $\vbfISO$, drawn from a Maxwellian distribution with zero mean and standard deviation $\sigISO\!\!=\!\!\{[1\!-\!\alfROT]$ $\Omega/6\Mtot\}^{1/2}$, where $\Mtot$ is the total mass of the subcluster. These velocities are then shifted to the centre-of-mass frame, and renormalised so that the total kinetic energy invested in isotropic velocity dispersion is correct.

If $\alfROT>0$, the stars are also given ordered rotation velocities, $\vbfROT$. If $\alfLAW=\mbox{\sc sol}$ the rotation is solid-body, with moment of inertia about the $z$-axis, angular speed about the $z$-axis, and velocity given respectively by
\begin{eqnarray}\label{EQN:SOL}
\left.\begin{array}{rcl}
\Iz&=&\sum\limits_{n=1}^{n=N}\{\Mn|\ebfhatz\!\wedge\!\rbfn|^2\}\!,\\
\omega&=&\left[\alfROT\Omega/\Iz\right]^{1/2}\!,\!\!\\
\vbfROTn&=&\omega\;\,\ebfhatz\!\wedge\!\rbfn\,,\\
\end{array}\right\}\;\mbox{if}\;\alfLAW\!=\mbox{\sc sol}\,;
\end{eqnarray}
here $\ebfhatz$ is the unit vector parallel to the $z$-axis and $\wedge$ denotes a vector product. Alternatively, if $\alfLAW=\mbox{\sc kep}$ the rotation is Keplerian, with gravitational potential relative to the $z$-axis, Kepler coefficient, and velocity given respectively by
\begin{eqnarray}\label{EQN:KEP}
\left.\begin{array}{rcl}
\Iz'&=&\sum\limits_{n=1}^{n=N}\{\Mn|\ebfhatz\!\wedge\!\rbfn|^{-1}\}\!,\\
\kappa\!\!&\!=\!&\!\!\left[\alfROT\Omega/\Iz'\right]^{1/2}\!,\!\!\\
\vbfROTn\!\!\!&\!=\!&\!\!\kappa\;|\ebfhatz\!\wedge\!\rbfn|^{-3/2}\;\ebfhatz\!\wedge\!\rbfn\!,\!\!\\
\end{array}\right\}\;\mbox{if}\;\alfLAW\!=\mbox{\sc kep}.\;
\end{eqnarray}
The net velocities of the stars,
\begin{eqnarray}
\vbfn&=&\vbfISOn\,+\,\vbfROTn,
\end{eqnarray}
are then computed and shifted (again) to the centre-of-mass frame.

Finally the initial total kinetic energy,
\begin{eqnarray}
\Ekin&=&\sum\limits_{n=1}^{n=N}\,\left\{\frac{\Mn\;|\vbfn|^2}{2}\right\}
\end{eqnarray}
is computed, and the velocities are re-scaled to ensure that the subcluster is virialised,
\begin{eqnarray}
\vbfn&\longleftarrow&\vbfn\;\left[-\,\Omega/2\Ekin\right]^{1/2}\,.
\end{eqnarray}
We invoke an initially virialised subcluster for two reasons. First, the timescale on which the subcluster approaches virial equilibrium is likely to be short, compared with the timescale on which the multiplicity statistics stabilise. Second, this reduces the number of free parameters we need to explore.

We introduce the shorthand
\begin{eqnarray}
{\cal C}&\equiv&[\sigell,\,\alfROT,\,\alfLAW,\,\alfSEG]\hspace{0.8cm}
\end{eqnarray}
to represent a specific set of Configuration Parameters. Apart from $N$ and ${\cal C}$, the only other variables that distinguish one experiment from another are the random-number seeds used to generate different realisations of the same $N$ and same set of Configuration Parameters, ${\cal C}$.

\subsection{Evolving Subclusters}

Subclusters are evolved using a 4th-order Runge-Kutta integration scheme with an adaptive timestep,
\begin{eqnarray}\label{EQNDeltat}
\Delta t&=&\gamma\;{\rm MIN}\!\left\{\frac{|\rbfn-\rbfnp|}{|\vbfn-\vbfnp|},\,\frac{|\vbfn-\vbfnp|}{|\abfn-\abfnp|}\right\}_{n\neq n'}.
\end{eqnarray}
Here $\gamma$ is a user-specified parameter determining the accuracy of the integration, $\abfn$ is the acceleration of star $n$, and the minimum is over all possible pairs of stars. This formulation ensures that no two stars change their relative position or relative velocity by more than a fraction $\sim\!\gamma$ in one timestep. The numerical experiments presented in the sequel have been performed with $\gamma\!=\!0.1$.

Each initial system is integrated for a maximum of 1000 crossing times ($\sim 70\,{\rm Myr}\,N^{-1/2}$), unless all the existing systems become unbound from one another, in which case the integration is terminated in order to save computer time (see Section \ref{unbsec}). Conservation of linear momentum, angular momentum and energy is monitored throughout the evolution.

\subsection{Identifying Multiple Systems}

At regular intervals, $\DtMMO$, during the cluster evolution, we identify multiple systems using a Multiplicity Monitoring Operation, hereafter MMO. The MMO selects a random pair of stars in the subcluster. If the stars in this pair [1] are mutual nearest neighbours, and [2] have negative energy in their mutual centre of mass frame, they are identified as bound. For the purposes of the MMO, the bound pair is then treated as a single object with the properties (mass, position, velocity) of its center of mass. This object is added to the pairing procedure, and its individual constituent stars are removed from consideration for the purpose of this  implementation of the MMO. Conversely, if the pair do not meet both criteria [1] and [2], they remain as individual stars, and the MMO checks a different possible pairing. This continues recursively until the full inventory of multiple systems in the subcluster has been determined.

For each identified bound system -- including higher-order multiples -- the procedure calculates the orbital parameters: semi-major axis ($a$), orbital period ($P$), eccentricity ($e$), and angular momentum vector (which gives the orbital inclination $\hat{\boldsymbol k}$). The MMO is implemented every 33 crossing times ($\sim2.3\,{\rm Myr}\,N^{-1/2}$, or $\sim1\,{\rm Myr}$ for $N=4\;{\rm or}\;5$). Note that, after an MMO implementation, the integration procedure continues to  follow all stars as individuals.

\subsection{Unbound Stars and the Stopping Condition} \label{unbsec}

During the evolution, a star or system may become unbound from the rest of the subcluster. It is then no longer necessary to track its position, and it is removed from further evolution of the subcluster (and from subsequent MMO implementations). However, during periods of frequent close encounters two stars that are actually bound to one another may not meet condition [1] (i.e. may briefly not be mutual nearest neighbours). This only occurs occasionally, and only ever in the very early lifetime of a subcluster when the dynamics is very chaotic. To avoid inadvertently removing bound stars or systems from the evolution, stars and systems are only removed after the first 200 crossing times, and only if those stars or systems have remained unbound for two consecutive MMOs.

If the MMO finds that the subcluster consists entirely of singles and binaries that are unbound from one another,  no further dynamical change can take place and the evolution is halted.

\subsection{Comparison with Observations} \label{mixsec}

For each set of Configuration Parameters, ${\cal C}$, and number of stars, $N$, we evolve 1000 different realisations and compute the mean number of systems with multiplicity $m$ formed per realisation, $\SCNm.$ We do this for all $N$ in $3\!\leq\!N\!\leq\!7$, for each set of Configuration Parameters, ${\cal C}$, as listed in Table 2, and for all $m$ in $1\!\leq\!m\!\leq\!\mmax\!=\!4^+$. Because the numbers of quintuples, sextuples and septuples are small (both in the numerical experiments, and in the observations), they are simply added to the quadruples to give $\SCNfp$.

To fit the observations, we assume that the probability that a core spawns a subcluster of $N$ stars is given by the distribution function
\begin{eqnarray}\label{EQN:PDN.01}
\PDN\!\!&\!\!=\!\!&\!\!\etD\,{\rm MAX}\!\left(\left\{\DND^2\,-\,\left[N-\ND\right]^2\right\},\;0\right),\hspace{0.5cm}\\\label{EQN:etD.01}
\etD\!\!&\!\!=\!\!&\!\!1\,\Big/\,\sum\limits_{N=1}^{N=7}\,{\rm MAX}\!\left(\left\{\DND^2\,-\,\left[N-\ND\right]^2\right\},\;0\right),\hspace{0.7cm}
\end{eqnarray}
where $\etD$ is a normalisation coefficient. Because $N$ is not a continuous variable, $\ND$ regulates, but is not exactly, the mean of the $N$ distribution; and $\DND$ regulates, but is not exactly, the standard deviation of the $N$ distribution. The true mean and standard deviation are given by
\begin{eqnarray}\label{EQN:muD.01}
\muD&=&\sum\limits_{N=1}^{N=7}\,\left\{\PDN\,N\right\},\\\label{EQN:sigmaN.01}
\sigD^2+\muD^2&=&\sum\limits_{N=1}^{N=7}\,\left\{\PDN\,N^2\right\}.\hspace{0.6cm}
\end{eqnarray}
The subscript ${\cal D}$ in Equations \ref{EQN:PDN.01} through \ref{EQN:sigmaN.01} represents the parameters of the distribution function, i.e.
\begin{eqnarray}\label{EQN:D.01}
{\cal D}&\equiv&\left[\ND,\,\DND\right],
\end{eqnarray}
hereafter the $N$-Distribution.

The frequency with which a subcluster having Configuration Parameters ${\cal C}$ and $N$-distribution ${\cal D}$ spawns a multiple system of order $m$ is
\begin{eqnarray}
\fCDm&=&\sum\limits_{N=1}^{N=7}\!\left\{\PDN\,\SCNm\right\}.
\end{eqnarray}
Therefore, if the observed number of systems with multiplicity $m$ is ${\cal O}_{_m}$, the predicted number is
\begin{eqnarray}
{\cal N}_{_{{\cal CD}m}}&=&f_{_{{\cal CD}m}}\,\sum\limits_{m=1}^{m=4^+}\!\left\{{\cal O}_{_m}\right\}\,\Big/\,\sum\limits_{m=1}^{m=4^+}\!\left\{f_{_{{\cal CD}m}}\right\},\hspace{0.5cm}
\end{eqnarray}
and the quality of fit for this combination of Configuration Parameters, ${\cal C}$, and $N$-Distribution, ${\cal D}$, is given by
\begin{eqnarray}
\QCD&=&\sum\limits_{m=1}^{m=4^+}\!\left\{\frac{[{\cal N}_{_{{\cal CD}m}}-{\cal O}_{_m}]^2}{{\cal O}_{_m}^2}\right\}
\end{eqnarray}
A lower value of $\QCD$ represents a better fit.

\section{Results and Discussion}\label{results}

\begin{figure*}
    \centering
    \includegraphics[width=11cm]{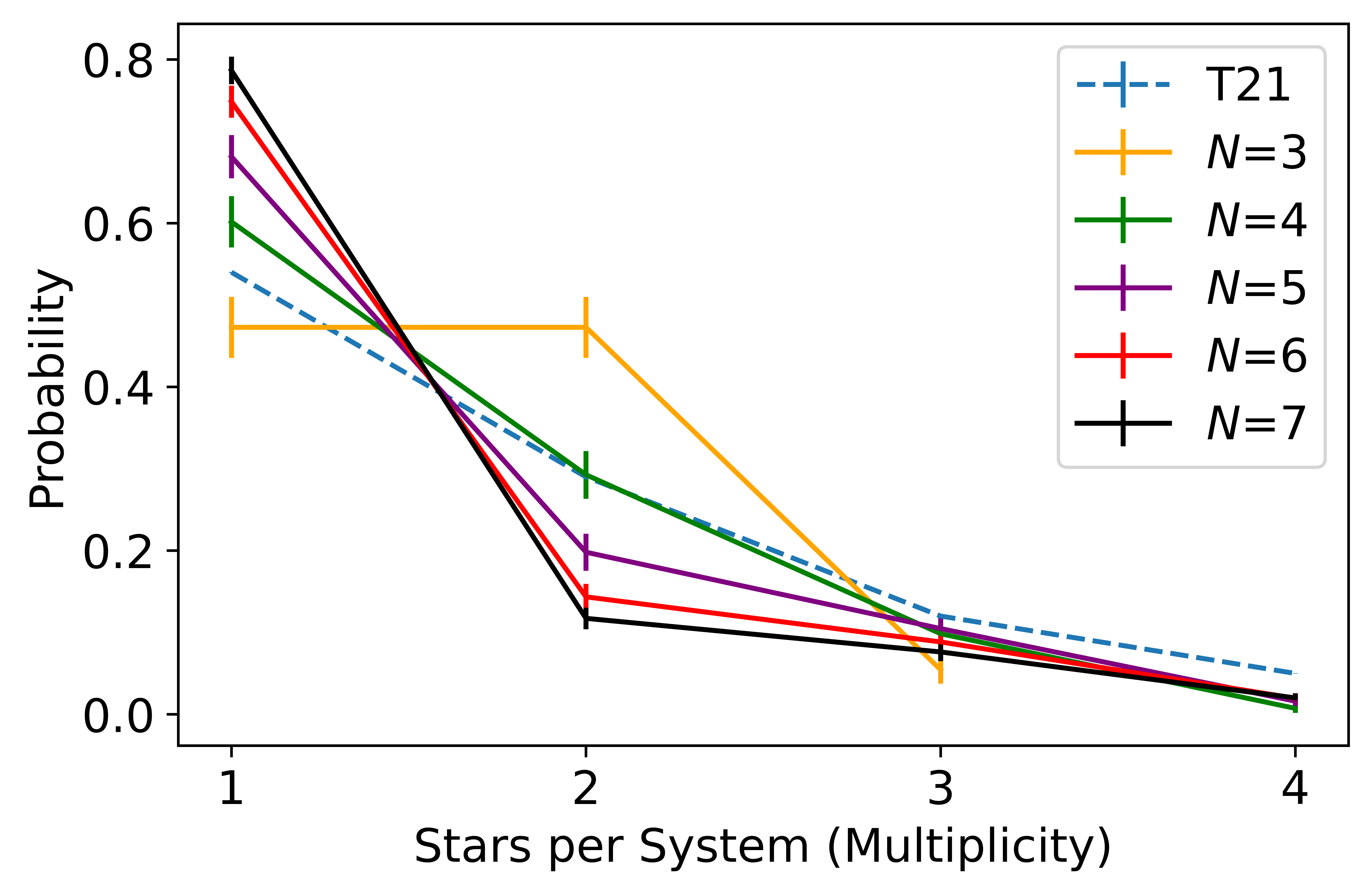}
    \caption{Multiplicity distributions for subclusters with $\,N\!=\!3$ (yellow), 4 (green), 5 (purple), 6 (red) and 7 (black). For each $N$-value we have evolved 1000 different realisations with the fiducial Configuration Parameters ($\alfROT =0$, $\alfSEG=0$). The blue dashed line represents the T21 sample. Error bars represent the 3$\sigma$ statistical variance.}
    \label{fig:MultN0}
\end{figure*}

\begin{figure*}
    \centering 
    \includegraphics[width=11cm]{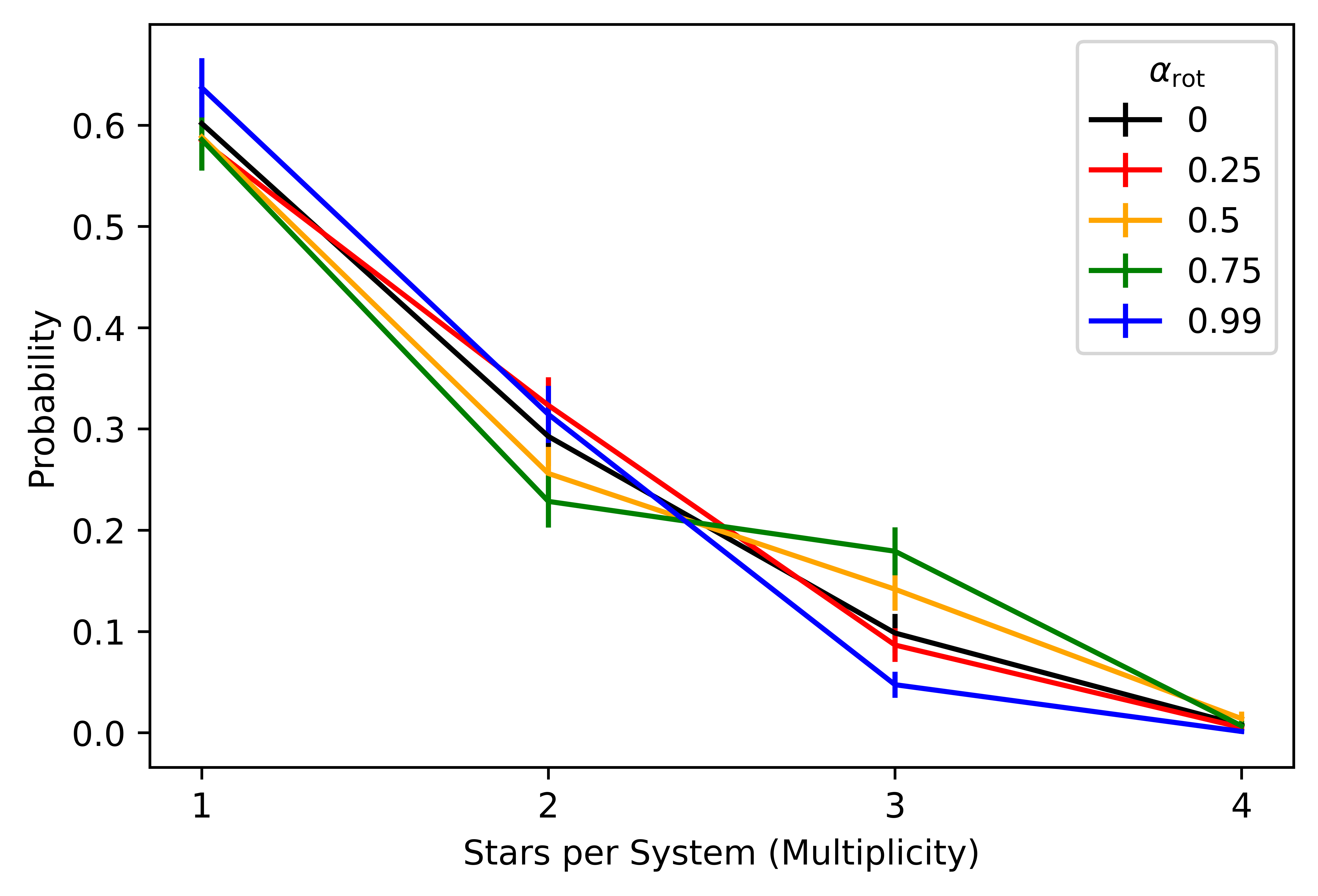}
    \caption{Multiplicity distributions for subclusters with $N\!=\!4$, $\alfSEG=1$, and different values of $\alfROT$ = 0 (black), 0.25 (red), 0.5 (yellow), 0.75 (green), and 0.99 (blue). For each $\alfROT$ value we have evolved 1000 different realisations. Error bars represent the 3$\sigma$ statistical variance.}
    \label{fig:MultBeta}
\end{figure*}

\begin{table*}
\caption{The Configurations evaluated, and the quality and parameters of the best fit. ~ Column 1, Configuration ID; Columns 2 through 4, the amount of rotation, the rotation law, and whether there is mass segregation; Columns 5 and 6, the N-Distribution Parameters, $(\ND,\DND)$; Columns 7 and 8, the mean and standard deviation of the $N$-Distribution; Column 9, the quality of the best fit (small for a good fit); Columns 10 through 13, the percentages of singles, S, binaries, B, triples, T, and quadruples plus higher orders, Q$^+$; Column 14, the total number of systems produced; Columns 15 through 18, the multiplicity fraction (MF, Equation \ref{EQN:MF}), the triple and higher-order fraction ($\fTp$, Equation \ref{EQN:fTp.01}), the companion fraction (CF, Equation \ref{EQN:CF}), and the plurality (PL, Equation \ref{EQN:PL}). ~ Row 1, the parameter symbols; Row 2, the observed statistics from the T21 sample \citep{2021Univ....7..352T}; Rows 3 through 5, the results from \citet{1998A&A...339...95S} for subclusters with a single $N=3$, 4 and 5; Rows 6 through 10, the results from this work for subclusters with a single $N=3$, 4, 5, 6 and 7; Row 11, the fiducial case (no rotation, no segregation); Rows 12 through 15, the solid-body rotation cases; Rows 16 and 17, the Keplerian rotation cases; Rows 18 through 22, the mass-segregated cases (including the best-fit case, SEG4).}

\begin{center} \label{table:results}
\begin{tabular}{cccccccccccccccccc}\hline
${\cal C}$ & $\alfROT$ & $\alfLAW$ & $\alfSEG$ & $N_{_{\cal D}}$ & $\Delta N_{_{\cal D}}$ & $\mu_{_D}$ & $\sigma_{_D}$ & $\QCD$ & S & B & T & Q$^+$ & Sys & MF & $\fTp$ & CF & PL \\\hline
T21 & 0 & -- & -- &  &  &  & &  & 54 & 29 & 12 & 5 &  & 0.46 & 0.17 & 0.68 & 1.13 \\
\hline
SD3 & 0 & -- & 0 & 3 & 0 &  & &  & 46.8 & 46.8 & 5.5 & 0 & 1,878 & 0.53 & 0.06 & 0.59 & 0.82 \\
SD4 & 0 & -- & 0 & 4 & 0 &  & &  & 62.7 & 29.1 & 6.0 & 1.1 & 2,713 & 0.36 & 0.07 & 0.44 & 0.74 \\
SD5 & 0 & -- & 0 & 5 & 0 &  & &  & 70.1 & 18.9 & 9.0 & 0.9 & 3,475 & 0.29 & 0.10 & 0.40 & 0.75 \\\hline
{\it N}=3 & 0 & -- & 0 & 3 & 0 & 3 & 0 & 1.71 & 47.3 & 47.3 & 5.45 & 0 & 1,614 & 0.53 & 0.05 & 0.58 & 0.80 \\
{\it N}=4 & 0 & -- & 0 & 4 & 0 & 4 & 0 & 0.77 & 60.2 & 29.3 & 9.85 & 0.73 & 2,194 & 0.40 & 0.11 & 0.51 & 0.84 \\
{\it N}=5 & 0 & -- & 0 & 5 & 0 & 5 & 0 & 0.65 & 68.1 & 19.8 & 10.5 & 1.61 & 2,798 & 0.32 & 0.12 & 0.46 & 0.84 \\
{\it N}=6 & 0 & -- & 0 & 6 & 0 & 6 & 0 & 0.84 & 74.8 & 14.3 & 8.84 & 1.96 & 4,377 & 0.25 & 0.11 & 0.38 & 0.76 \\
{\it N}=7 & 0 & -- & 0 & 7 & 0 & 7 & 0 & 1.06 & 78.7 & 11.7 & 7.62 & 2.00 & 5,290 & 0.21 & 0.10 & 0.33 & 0.70 \\\hline
{\sc FID} & 0 & -- & 0 & 4.9 & 1.0 & 4.8 & 0.1 & 0.64 & 67.1 & 21.0 & 10.4 & 1.49 & 16,273 & 0.33 & 0.12 & 0.46 & 0.84 \\\hline
SOL1 & 0.25 & {\sc sol} & 0 & 5.0 & 1.0 & 5.0 & 0.0 & 0.57 & 66.2 & 21.7 & 10.4 & 1.7 & 17,104 & 0.34 & 0.12 & 0.48 & 0.86 \\
SOL2 & 0.50 & {\sc sol} & 0 & 4.3 & 3.3 & 4.4 &  2.1 &  0.41 & 62.6 & 24.3 & 11.1 & 2.0 & 16,599 & 0.37 & 0.13 & 0.53 & 0.91 \\
SOL3 & 0.75 & {\sc sol} & 0 & 4.4 & 3.4 & 4.4 & 2.2 & 0.49 & 63.8 & 23.1 & 11.3 & 1.78 & 16.714 & 0.36 & 0.13 & 0.51 & 0.90 \\
SOL4 & 0.99 & {\sc sol} & 0 & 4.9 & 1.0 & 4.8 & 0.1 & 1.26 & 70.3 & 23.7 & 5.65 & 0.38 & 17,829 & 0.30 & 0.06 & 0.36 & 0.63 \\\hline
KEP1 & 0.50 & {\sc kep} & 0 & 4.4 & 3.4 & 4.4 & 2.2 & 0.42 & 63.5 & 23.6 & 10.8 & 2.08 & 16,715 & 0.36 & 0.13 & 0.51 & 0.90 \\
KEP2 & 0.99 & {\sc kep} & 0 & 5.0 & 1.0 & 5.0 & 0.0 & 1.22 & 71.2 & 21.8 & 6.57 & 0.38 & 17,992 & 0.29 & 0.07 & 0.36 & 0.64 \\\hline
SEG1 & 0 & -- & 1 & 5.0 & 1.0 & 5.0 & 0.0 & 0.63 & 67.1 & 20.6 & 10.8 & 1.53 & 16,590 & 0.33 & 0.12 & 0.47 & 0.85 \\
SEG2 & 0.50 & {\sc sol} & 1 & 5.3 & 1.6 & 5.2 & 0.5 & 0.43 & 66.1 & 21.0 & 10.6 & 2.29 & 16,574 & 0.34 & 0.13 & 0.49 & 0.89 \\
SEG3 & 0.99 & {\sc sol} & 1 & 4.6 & 1.1 & 4.6 & 0.2 & 1.38 & 69.5 & 25.9 & 4.31 & 0.33 & 17,869 & 0.31 & 0.05 & 0.36 & 0.60 \\
SEG4 & 0.50 & {\sc kep} & 1 & 5.4 & 4.4 & 4.8 & 2.4 & 0.35 & 63.2 & 24.1 & 10.3 & 2.41 & 16,339 & 0.27 & 0.13 & 0.52 & 0.91 \\
SEG5 & 0.99 & {\sc kep} & 1 & 4.8 & 1.0 & 4.7 & 0.2 & 1.33 & 69.9 & 25.1 & 4.59 & 0.41 & 17,995 & 0.30 & 0.05 & 0.36 & 0.61 \\\hline 
\end{tabular}
\end{center}
\label{TAB:Configurations}
\end{table*}

\begin{figure*}
    \centering
    \includegraphics[width=11cm]{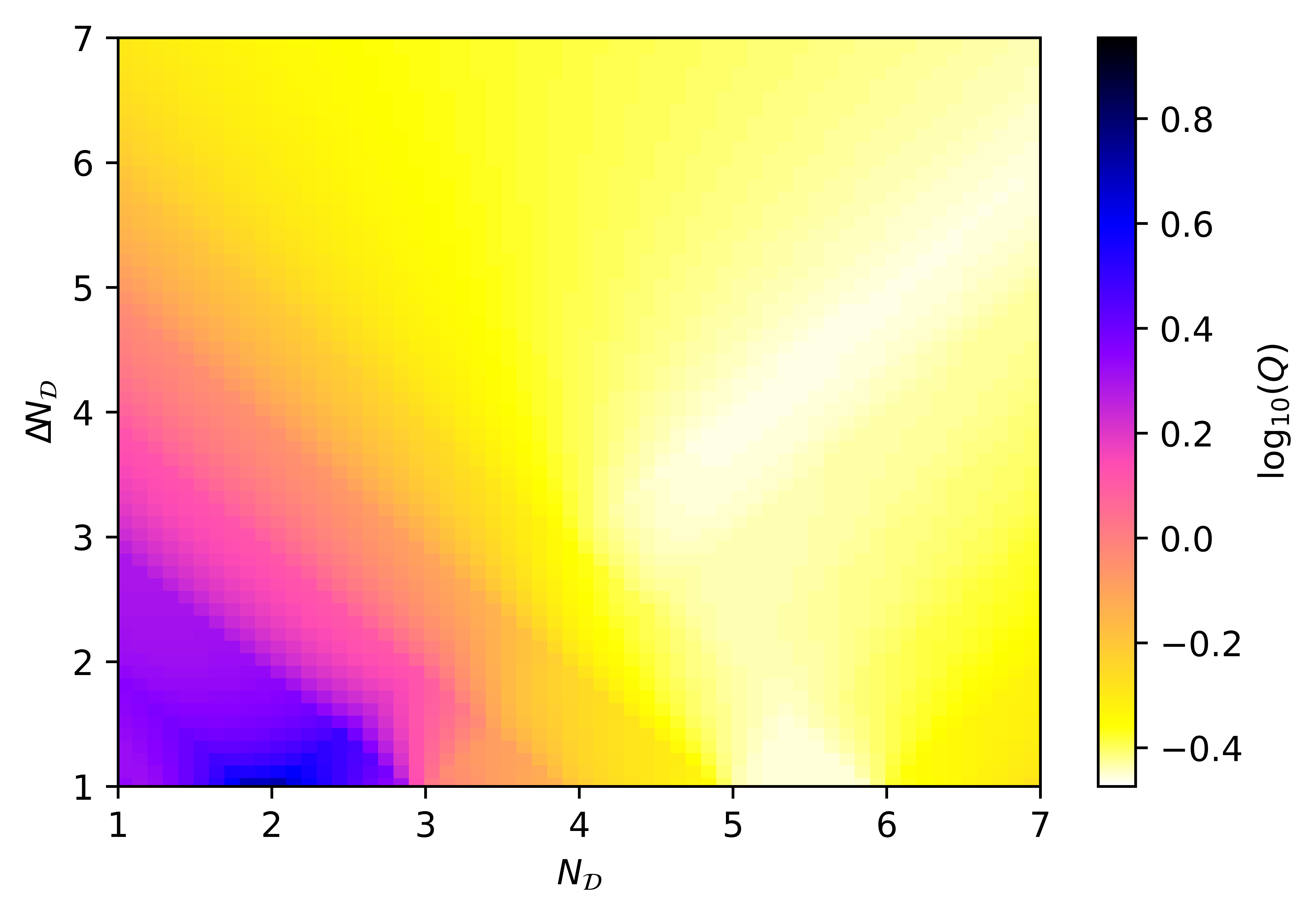}
    \caption{Map of the Quality Factor, $\QCD$, for  the best-fit Configuration Parameters ($\alfROT=0.50$, $\alfLAW=$ KEP, $\alfSEG=1$) and the full range of $N$-Distribution Parameters ($1.0\leq\ND\leq7.0$ and $1.0\leq\DND\leq7.0$). The colour encodes $\log_{_{10}}\!(\QCD)$, with the best fits white, very bad fits purple, and the worst fit ($\ND=2$ and $\DND=1$) black.}
    \label{fig:QTok}
\end{figure*}

\begin{figure*}
    \centering
    \includegraphics[width=11cm]{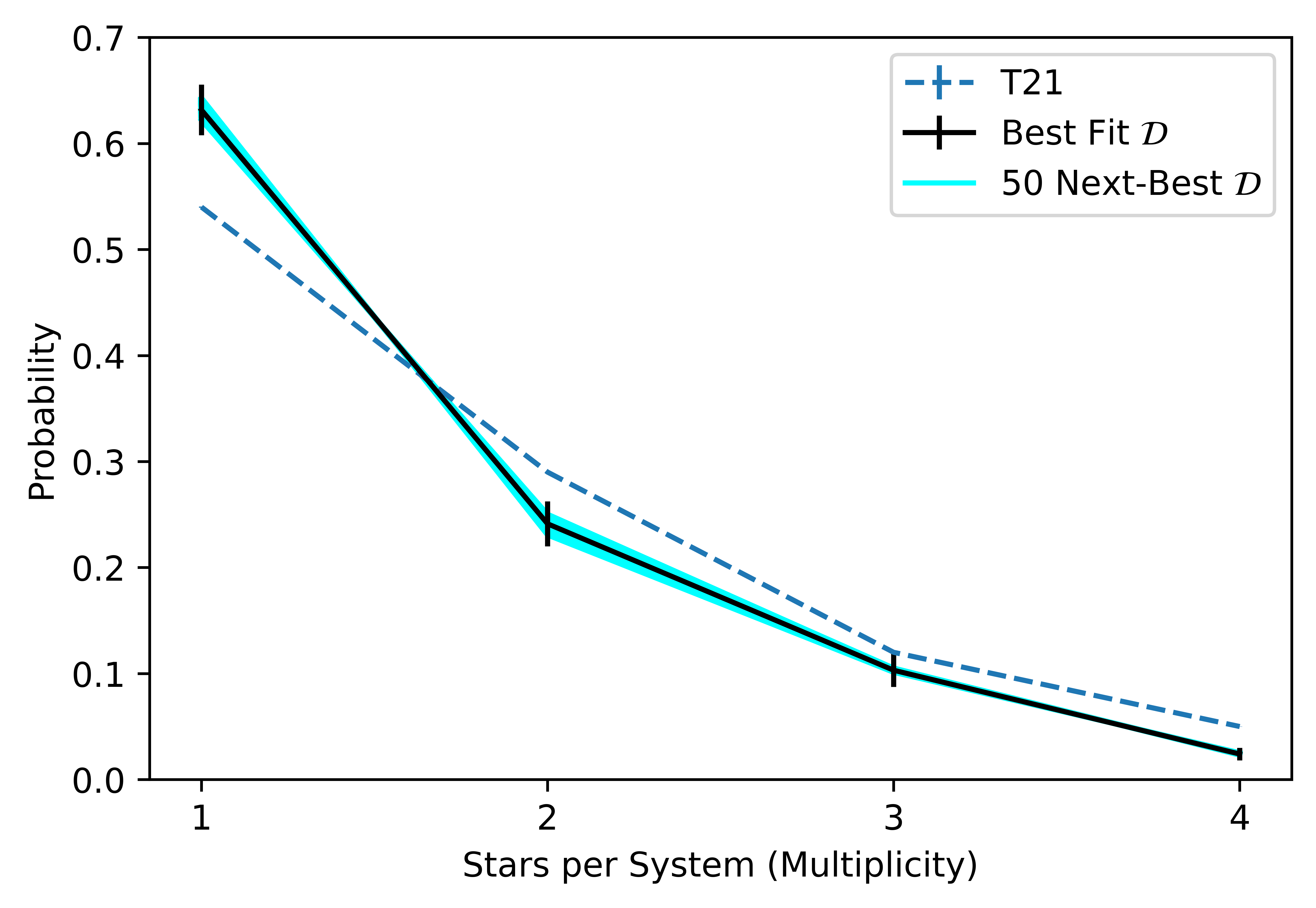}
    \caption{Multiplicity distributions. The blue dashed curve represents the T21 sample. The black full line represents the absolute best-fit combination of Configuration Parameters and $N$-Distribution Parameters (i.e. SEG4, one from bottom row in Table \ref{TAB:Configurations}). The surrounding cyan represents fits for the same Configuration Parameters and the 50 next-best-fits with different $N$-Distribution Parameters (corresponding to the whitest region on Fig. \ref{fig:QTok}).}
    \label{fig:MultTok}
\end{figure*}

\begin{figure*}
    \centering
    \includegraphics[width=16cm]{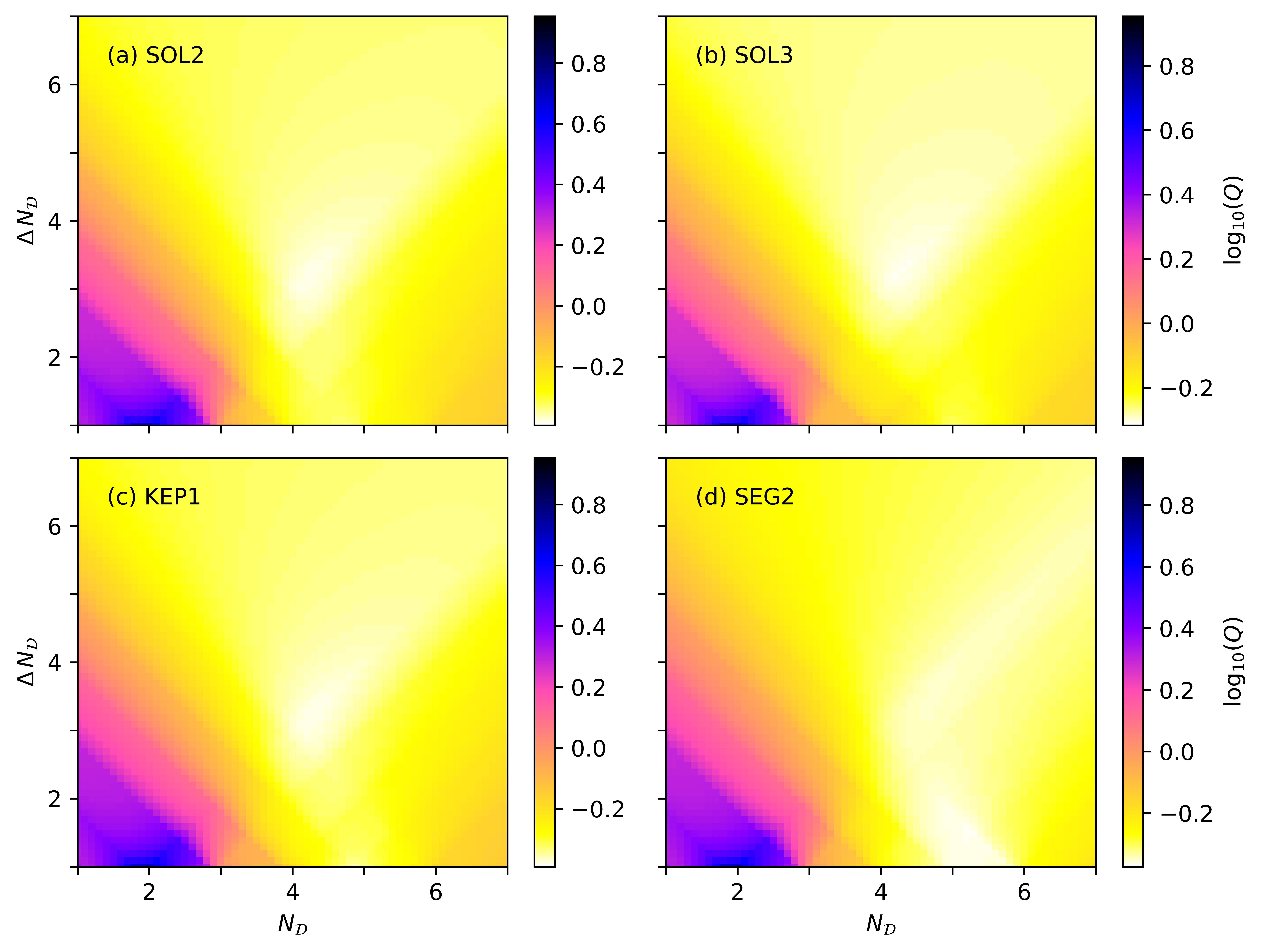}
    \caption{As Fig. \ref{fig:QTok}, but for the Configuration Parameters giving the second through fifth best fits (i.e. the other fits with $\QCD\!<\!0.50$): (a) SOL2 with $\QCD\!=\!0.41$; (b) SOL3 with $\QCD\!=\!0.49$; (c) KEP1 with $\QCD\!=\!0.42$; and (d) SEG2 with $\QCD\!=\!0.43$.}
    \label{fig:QMon}
\end{figure*}

Our numerical experiments suggest that most cores spawn 4 or 5 stars. There are cores that spawn 3 stars, but if they were the norm they would produce too many binaries and too few singles, while contributing no quadruples or higher-order systems. Conversely, there are cores that spawn more than 5 stars, but if they were the norm they would produce too many singles and too few binaries. Although such high-$N$ cores are essential to produce the high-order systems, our pure $N$-body model does not produce as many high-order systems as are observed; we find that increasing $N$ is more effective at over-producing singles than it is at producing stable higher-order multiples. In Section \ref{SEC:Conclusions} we suggest that forming the observed number of high-order systems probably requires the inclusion of additional physics such as disc drag and disc fragmentation, since this will deliver compact, tightly bound companions that can survive all but the closest interactions with other stars.

\subsection{A universal {\it N} value}\label{SEC:UniversalN}

It is informative to consider the possibility that all cores spawn the same number of stars, i.e. a universal $N$ value, although a universal N is extremely unlikely. For these experiments we use the fiducial parameters, i.e. no rotation and no mass-segregation.

An initially-bound subcluster with $\,N\!=\!3\,$ stars has $\,{\cal A}_{_3}=2\,$ possible end-states: (i) a triple system, or (ii) a binary and a single star. Because the dissolution of a 3-star subcluster into a binary and a single is the only way for a binary or a single star to form from such a subcluster, the percentages of single and binary systems must be equal in this case. For the fiducial case, in which the initial subcluster has no ordered rotation or mass segregation, $\sim90\%$ of 3-star subclusters decay into a binary and a single, giving percentages of S:B:T:Q$^+$ = 47:47:6:0, exactly as obtained by \cite{1998A&A...339...95S}.

An initially-bound subcluster with $\,N\!=\!4\,$ stars has $\,{\cal A}_{_4}=4\,$ possible end-states: (i) a quadruple; (ii) a triple and a single star; (iii) two separate binaries; or (iv) a binary and two single stars. For the fiducial case, $\,N\!=\!4\,$ gives S:B:T:Q$^+$ = 60:29:10:7. We note that higher-order multiples may come in different variants. For example, there are `2+2' quadruples and planetary quadruples.

Of the quadruple systems produced in the $\,N\! = \!4\,$ fiductial, $25(\pm 14)\%$ are `2+2' systems and the remainder are planetary. In contrast, in the \citealt{2014AJ....147...87T} sample $\sim75\%$ of quadruples are `2+2' systems. There are two possible explanations for this discrepancy. First, it is probably easier to detect `2+2' quadruples spectroscopically, in which case they are over-represented in the observational sample. Second, it may be that additional physics that is missing from our numerical experiments -- for example, circumstellar accretion discs and disc fragmentation -- increases the percentage of `2+2' quadruples.

The number of distinguishable end-states increases monotonically with $\,N.\,$ 5-star subclusters have $\,{\cal A}_{_5}=6\,$ possible end-states, and 6-star subclusters have $\,{\cal A}_{_6}=10\,$ possible end-states. For the fiducial case, $\,N\!=\!5\,$ gives S:B:T:Q$^+$ = 68:20:10:2, with $24(\pm8)\%$ of Q$^+$ systems composed wholly, or in part, of `2+2' configurations; and $\,N\!=\!6\,$ gives S:B:T:Q$^+$ = 75:14:9:2, with $24(\pm6)\%$ of Q$^+$ systems containing `2+2' configurations.

An initially-bound subcluster with $\,N\!=\!7\,$ stars has $\,{\cal A}_{_7}=14\,$ possible end-states: (i) a septuple; (ii) a sextuple and a single star, (iii) a quintuple and a binary, (iv) a quintuple and two singles, (v) a quadruple and a triple, (vi) a quadruple, a binary, and a single, (vii) a quadruple and three singles, (viii) two triples and a single, (ix) a triple and two binaries, (x) a triple, a binary, and two singles, (xi) a triple and four singles, (xii) three binaries and a single, (xiii) two binaries and three singles, (xiv) a binary and five singles. Given the rapid increase in the number of end-states with increasing $N$ (see Equation \ref{EQN:decay} and Appendix \ref{appendA1}), we limit consideration to $\,N\!\leq\!7.\,$ For the fiducial case, $\,N\!=\!7\,$ gives S:B:T:Q$^+$ = 79:12:8:2, with $19(\pm 5)\%$ of Q$^+$ systems containing `2+2' configurations.

Figure \ref{fig:MultN0} displays the multiplicity distributions for the fiducial case with different values of $\,N.\,$ The results for $\,N\!=\!4\,$ most accurately match the T21 sample, but no value reproduces exactly the observed multiplicities. Increasing $N$ does increase the probability of forming higher-order multiples. Indeed, a system with multiplicity $m\!\geq\!3$ can only form if $N\!\geq\!m$. However, this effect is small, and the main effect of increasing $N$ is to increase the percentage of singles at the expense of binaries. These singles are mainly lower-mass stars that have been ejected by the sling-shot mechanism.

For subclusters with a universal $N$ value the multiplicity statistics can be changed quite significantly by introducing rotation. Figure \ref{fig:MultBeta} shows the effect of different amounts of rotation for subclusters with $N\!=\!4$.

\subsection{A distribution of {\it N} values}\label{SEC:DistributedN}

It is unlikely that every prestellar core produces the same number of stars. Therefore we now consider $N$-Distributions, and identify the combinations of Configuration Parameters, ${\cal C}$, and $N$-Distribution, ${\cal D}$, that best fit the observations. The $N$-Distributions are given by Equation \ref{EQN:PDN.01}, and we test all combinations of $\,N_{_{\cal D}}\,=$ 1.0, 1.1, 1.2, 1.3,  . . . 6.8, 6.9, 7.0, and $\,\Delta N_{_{\cal D}} =$ 1.0, 1.1, 1.2, 1.3,  . . . 6.8, 6.9, 7.0.

Table \ref{table:results} lists all the tested Configurations ${\cal C}$; for each Configuration, the parameters of the $N$-Distribution giving the best fit to the T21 sample; and -- where they exist -- the corresponding parameters from  \citet{1998A&A...339...95S}.

\subsubsection{The notional absolute-best fit}

The notional absolute-best fit is obtained with Configuration ${\cal C}=$ SEG4 ($\alfROT =0.5$, $\alfLAW=\mbox{\sc kep}$, $\alfSEG=1$) and an $N$-Distribution with $(\ND,\DND)\!\simeq\!(5.4,4.4)$ (corresponding to mean $\muD\simeq4.8$, and standard deviation $\sigD\simeq2.4$). This delivers percentages S:B:T:Q$^+=63:24:10:2.4$, with Quality Factor $\QCD=0.35$ and  $41(\pm 5)\%$ of Q$^+$ systems containing `2+2' orbits.

\subsubsection{How critical is the $N$-Distribution?}

Figure \ref{fig:QTok} is a map of the Quality Factor, $\QCD$, for Configuration ${\cal C}=$ SEG4 over the full range of $N$-Distribution parameters, $(\ND,\DND)$. There are two regions of low $\QCD$ (i.e. good fits, represented by white on Fig. \ref{fig:QTok}). One is centred on the best fit $(\ND,\DND)\!\simeq\!(5.4,4.4)$ with an extension to higher and lower values of $\ND$ and $\DND$. The other, slightly less-favoured region, is centred on $(\ND,\DND)\!\simeq\!(5.5,1)$.

Figure \ref{fig:MultTok} compares the multiplicities corresponding to the absolute-best fit (black line) and the fifty next-best fits (i.e. SEG4 with slightly different $N$-Distributions; cyan shading). These fits correspond to the whitest region on Figure \ref{fig:QTok} (specifically, an area that is $\sim\!1.4\%$ of the total area of Fig. \ref{fig:QTok}). They all fall within one standard deviation of the absolute-best fit, so the parameters of the $N$-Distribution giving the absolute-best fit are not highly critical.

Notably, but unsurprisingly, a prestellar core producing exactly 2 stars results in the worst fit, since it only produces binary systems. The corresponding black point on Figure \ref{fig:QTok} is only just visible next to the abscissa.

\subsubsection{Which are the critical Configuration Parameters?}

Table \ref{TAB:BestFit} lists the five Configurations that produce fits with, $\QCD <0.5$.

\begin{table}
\caption{The parameters of the five best fits, i.e. those with Quality Factor, $\QCD<0.5$. Reading left to right, the columns give the configuration name, the parameters of the configuration ($\alfROT$, $\alfLAW$, $\alfSEG$), the mean and standard deviation of the distribution of $N$ values ($\muD$, $\sigD$), the Quality Factor ($\QCD$), and the percentage of Q$^+$ systems that contain `2+2' orbits.}
\begin{center}
\begin{tabular}{cccccccc}
${\cal C}$ & $\alfROT$ & $\alfLAW$ & $\alfSEG$ & $\muD$ & $\sigD$ & $\QCD$ & `2+2' \\
SOL2 & 0.50 & {\sc sol} & 0 & 4.4 & 2.1 & 0.41 &  $25(\pm 06)$ \\ 
SOL3 & 0.75 & {\sc sol} & 0 & 4.4 & 2.2 & 0.49 &  $26(\pm 06)$ \\ 
KEP1 & 0.50 & {\sc kep} & 0 & 4.4 & 2.2 & 0.42 &  $28(\pm 05)$ \\ 
SEG2 & 0.50 & {\sc sol} & 1 & 5.2 & 0.5 & 0.43 &  $44(\pm 07)$ \\ 
SEG4 & 0.50 & {\sc kep} & 1 & 4.8 & 2.4 & 0.35 &  $41(\pm 05)$ \\ 
\end{tabular}
\end{center}
\label{TAB:BestFit}
\end{table}

Figure \ref{fig:QMon} shows the Quality Factor, $\QCD$, for the four additional Configurations and the full range of $N$-Distribution parameters, $(\ND,\DND)$. Panels (a) SOL2, (b) SOL3, and (c) KEP1 on Fig. \ref{fig:QMon} are very similar to one another, with a single region of high quality, centred on $(\ND,\DND)\!\simeq\!(4.4,3.4)$. Panel (d) SEG2 is more like SEG4 (Figure \ref{fig:QTok}), with two regions, but the preferred region is now the lower one centred on $(\ND,\DND)\!=\!(5.3,1.6)$.

All five Configurations involve rotation, four with 50\% of the kinetic energy invested in rotation, and one with 75\%. We conclude that having comparable amounts of energy in rotation and in random isotropic motions is a critical requirement for producing a good fit.

In contrast, three of the top-five Configurations have solid-body rotation, whilst two have Keplerian rotation. We conclude that the details of the rotation (the rotation law) are not critical. Similarly three of the top-five configurations have no mass segregation, and two of them do. Therefore it appears that mass segregation is also not a critical requirement for a good fit.

The mass segregation does appear to have an affect upon the formation of '2+2' systems. While the percentage of '2+2's is similar for the best-fit cases with no segregation ($\alpha _{\rm seg}=0$), those that begin with segregated masses ($\alpha _{\rm seg}=1$) have a much higher percentage of '2+2's (Table \ref{TAB:BestFit}).

\subsection{Metrics of overall multiplicity}

Various metrics of overall multiplicity have been proposed, in particular the \textit{multiplicity fraction} (i.e. the fraction of systems that are not single),
\begin{eqnarray}\label{EQN:MF}
    \rm MF&=&\rm\frac{B+T+Q+...}{S+B+T+Q+...}\,;
\end{eqnarray}
the \textit{triple/higher-order fraction} (i.e. the fraction of systems that are triple or higher-order, see Equation \ref{EQN:fTp.01}); the \textit{companion fraction} (i.e. the mean number of companions per primary),
\begin{eqnarray}\label{EQN:CF}
\rm CF&=&\rm\frac{B+2T+3Q+...}{S+B+T+Q+...}\,;
\end{eqnarray}
and the {\it plurality} (i.e. the mean number of companions per star, irrespective of whether it is a primary star),
\begin{eqnarray}\label{EQN:PL}
\rm PL&=&\rm\frac{2B+6T+12Q+...}{S+2B+3T+4Q+...}\,.
\end{eqnarray}
These metrics are given in the last four columns of Table \ref{table:results}. The last one, PL, has the merit that it has a clear physical meaning and reflects, more strongly than the others, the percentage of higher-order multiples. CF and PL can both be greater than one. Indeed, for the T21 sample, PL $=1.13$.

\subsection{Caveats}\label{caveats}

\subsubsection{Pure $N$-body dynamics}

The numerical experiments reported here involve pure $N$-body dynamics. Consequently -- modullo numerical errors --  the resulting multiplicities are determined solely by the initial conditions and gravitational interactions between point masses. Phenomena such as dissipation from the cloud or disc, which harden systems against dissolution and act to increase stellar multiplicity, are not considered. We expect the pure $N$-body results of this work to represent a conservative (i.e. low) estimate of the multiplicity which can be achieved from a subcluster of $N$ stars.

\subsubsection{Isolated subclusters}

Each subcluster is evolved in isolation. The subcluster cannot capture outside stars, nor can it be perturbed by stellar flybys. For low-$N$ subclusters, capture might increase the multiplicity metrics (e.g. MF, CF and PL), but for high-$N$ subclusters it would probably reduce the multiplicity metrics. Perturbations by stellar flybys would be likely to reduce the multiplicity metrics, for example by disrupting hierarchical triples.

\subsubsection{Duration of integration}\label{SEC:Duration}

Subclusters are evolved for a maximum of 1000 crossing times. In practice most of the final multiple systems are established early in the evolution, within the first 200 crossing times. For the $N=7$ case, whose systems take the longest to settle into their end states, more than 80\% of instantiations achieve their final $1000\,t_{\rm cross}$ multiplicities by $200\,t_{\rm cross}$, and more than 95\% by $600\,t_{\rm cross}$.

\subsubsection{Limited number of stars in subcluster}

We do not consider subclusters with $N>7$. As we increase $N$, the multiplicity distribution changes at a decreasing rate (see Figure \ref{fig:MultN0}): the percentages of binaries and triples decrease slightly, the percentage of singles increases, and the percentages of higher-order multiples ($m\!\geq\!4$) increase imperceptibly. For example, in the fiducial case, the multiplicity distributions for $N=6$ and 7 agree within their 3$\sigma$ uncertainties (Figure \ref{fig:MultN0}). This is true for all configurations tested. We expect multiplicities for values of $N>7$ to follow this trend, remaining very similar to the multiplicity values for $N=7$.

In addition, the complexity of possible end states increases dramatically with $N$ (see Appendix \ref{appendA1}), making higher-$N$ numerical experiments prohibitive from a book-keeping standpoint.

\subsubsection{Observational statistics}

The T21 observational statistics are probably influenced by selection effects. The likelihood is that, as new and improved techniques and strategies are developed, the multiplicities of the systems in the T21 sample will change, and there will probably be an increase in the proportion of higher-order systems. As the completeness of multiplicity surveys improves, the analysis presented in this work can easily be reapplied to updated observational statistics. However, the inability of our pure $N$-body model to produce enough higher-order systems is likely to remain a limitation.

\section{Conclusion}\label{conclusion}\label{SEC:Conclusions}

We have used $N$-body numerical experiments to determine the multiplicity statistics that result from small-$N$ subclusters of stars that interact only through their mutual gravity. These statistics have then been compared with the T21 sample of nearby systems with solar-mass primaries \citep{2021Univ....7..352T}. The subclusters are presumed to be the product of collapse and fragmentation in a single isolated prestellar core.

To produce an acceptable fit to the observed statistics, prestellar cores must -- on average -- spawn between 4.3 and 5.2 stars. This seems to be a rather compelling conclusion, in the sense that firstly it produces by far the best fit to the observations, and secondly a very similar conclusion has been drawn by two other, completely independent studies, {\it viz}. [1] \citet{2013MNRAS.432.3534H} using statistical arguments, and [2] \citet{LomaxOetal2015a} using SPH simulations. We stress that neither [1] nor [2] involves $N$-body numerical experiments, and they are therefore totally independent of the results reported here.

In the numerical experiments reported in this work, subclusters which begin with roughly half their kinetic energy invested in rotation produce the best fits to the T21 sample. Furthermore there is a broad range of setups that produce very similar fits, but they all have roughly half their kinetic energy invested in rotation. These setups do occasionally spawn fewer than four stars, or more than five, but this is relatively rare. The multiplicity statistics appear to be independent of whether the subcluster starts with a solid-body or Keplerian rotation law, and of whether the masses are initially segregated. The orbital statistics of the systems formed in this paper will be presented in a companion paper.

Although the overall fits obtained here are quite good, there are always too few systems with multiplicity $m\!\geq\!4$. We believe that this is because our stars do not have circumstellar discs. Such discs will make close encounters between stars dissipative, thereby increasing the formation of tight orbits and higher-order multiples. Such discs may also fragment to form close companions, some of which will survive interactions with other stars in the subcluster, and again this will promote the formation of higher-order multiples. This refinement will be explored in a future paper.

\section{Acknowledgements}

Hannah Ambrose is grateful for the support of an STFC doctoral training grant.

\section{Data Availability}

The data underlying this article will be shared on reasonable request
to the corresponding author.

\bibliographystyle{mnras}
\bibliography{bibl.bib}

\appendix\section{Enumerating End States} \label{appendA1}

An initially virialised subcluster of $N\geq2$ stars must produce at least one bound multiple system. This component contains $n$ stars, where $2\!<\!n\!\leq\!N$, and we can assume without loss of generality that $n$ is the largest or equal-largest multiplicity of the end state. In the case where this component is the only multiple, all remaining $N\!-\!n$ stars are single.

The remaining $N\!-\!n$ stars may also form bound systems. These systems may have any multiplicity $n'$, where $2\!<\!n'\leq\!n$. The number of additional component systems of multiplicity $n'$ for an existing component of multiplicity $n$ is given by the number of times $n'$ may be divided wholly into the number of remaining stars. So, end states containing a component system of multiplicity $n$ and a component system (or systems) of multiplicity $n'$ are counted with the following equation:
\begin{eqnarray}
{\cal{A}}_{_{N\leq7}}&=&\sum_{n=2}^N\,\left\{1+\sum_{n'=2}^{n}\floor*{\frac{N-n}{n'}}\right\}.
\end{eqnarray}
This equation will count all possible end states for initially virialised subclusters with $2<N\leq 7$.

For $7<N\leq 10$, it is possible to achieve end states with component systems of multiplicity $n,n'$, and $n''$, where $n''<n'$, i.e. $N=8$ may end in two separate triples and a binary, $N=10$ may result in a quadruple, a triple, and a binary. To count these states, we must add an additional level of recursion. For each set of $n,n',$ there are $N-(n+n')$ remaining stars. If $N-(n+n')>2$, the stars may form a multiple of order $2\leq n''<n'$. $n''=n'$ systems are counted in the previous sum. The number of components of order $n''$ is then given by
\begin{eqnarray}
n_{n''}&=&\floor*{\frac{N-(n+n')}{n''}}
\end{eqnarray}
and the full sum becomes Equation \ref{EQN:decay},
\begin{eqnarray*}\nonumber\
    \AN\!\!&\!\!\!=\!\!&\!\!\sum_{n=2}^{N}\!\left\{ 1+\sum_{n'=2}^{n}\!\left\{\vphantom{\sum_{n''=2}^{n'-1}}\!\floor*{\frac{N\!-\!n}{n'}}\,+\,H(N\!-\!(n+n'+2))\;\;\right.\right. \\
   \!\!&\!\!\!\!\!&\!\!\hspace{3.0cm}\left.\left.\sum_{n''=2}^{n'-1}\floor*{\frac{N-(n+n')}{n''}}\right\}\right\}.
\end{eqnarray*}

where $H$ represents the Heaviside function, i.e.
\begin{eqnarray}
H(h)\;\;=\;\;
\begin{cases}
\;1 , \hspace{0.5cm} \mbox{if}\;\;h\geq 0\,;\\
\;0 , \hspace{0.5cm} \mbox{if}\;\;h<0\,.\\
\end{cases}
\end{eqnarray}
The sum gives the multiplicity of each end component, and differentiates components only on multiplicity, not on the type of system (e.g. there is no discrimination between a quadruple which is planetary versus one which is a 2+2) or on which specific stars occupy which remnant system. Further terms may be added to count end states of $N\leq 10$.

\section{Scaling relations} \label{appendA2}

The numerical experiments described in the paper are strictly speaking dimensionless, and have only been scaled to $M_{_{\rm med}}\!=\!0.25\,\Msun\;(\mu_{_\ell}\!=\!-0.6)$ and $R=10^3\,{\rm AU}$ for the purpose of illustration. To scale a given experiment to a subcluster with a different total mass, $M'_{_{\rm tot}}$, and/or a different radius, $R'$, we must multiply all stellar and system masses by $f_{_M}\!=\!M'_{_{\rm tot}}/M_{_{\rm TOT}}$; all position vectors and orbital axes by $f_{_R}\!=\!R'/R$; the time and all orbital periods by $[f_{_R}^3/f_{_M}]^{1/2}$; and all velocities by $[f_{_M}/f_{_R}]^{1/2}$. Orbital eccentricities and inclinations are unchanged.

\end{document}